\documentclass[twocolumn,pra,10pt,notitlepage,showpacs,a4paper]{revtex4}
\usepackage[usenames]{color} %used for font color
\usepackage{amssymb} %maths
\usepackage{amsmath} %maths
\usepackage{braket}
\usepackage[utf8]{inputenc} %useful to type directly diacritic characters
\usepackage{cleveref}
\usepackage{graphicx}% include figure files
\usepackage{epstopdf}
\usepackage{dcolumn}% align table columns on decimal point
\usepackage{bm}% bold math 
\usepackage{siunitx}

\def\gappeq{\mathrel{ \rlap{\raise.5ex\hbox{$>$}}
                      {\lower.5ex\hbox{$\sim$}}  } }
\def\lappeq{\mathrel{ \rlap{\raise.5ex\hbox{$<$}}
                      {\lower.5ex\hbox{$\sim$}}  } }

\begin{document}

\preprint{PRA}

\title{Interferometry using Adiabatic Passage in Dilute Gas Bose-Einstein Condensates.}

\author{M. Rab$^1$, A. L. C. Hayward$^1$, J.H. Cole$^2$, A.D. Greentree$^{1,3}$, and A.M. Martin$^1$}

\affiliation{$^1$School of Physics, The University of Melbourne, Victoria 3010, Australia.\\ 
  $^2$Chemical and Quantum Physics, School of Applied Sciences, RMIT University, Victoria 3001, Australia.\\
  $^3$Applied Physics, School of Applied Sciences, RMIT University, Victoria 3001, Australia.}

\date{\today}

\begin{abstract}
  We theoretically examine three-well interferometry in Bose-Einstein condensates using adiabatic passage.
  Specifically, we demonstrate that a fractional coherent transport adiabatic passage protocol enables stable spatial
  splitting in the presence of nonlinear interactions. A reversal of this protocol produces a coherent recombination of
  the BEC with a phase-dependent population of the three wells. The effect of nonlinear interactions on the interferometric
  measurement is quantified and found to lead to an enhancement in sensitivity for moderate interaction strengths.
\end{abstract}
     
 \maketitle

 % 03.75.-b, Matter Waves
 % 07.77.-n, Atomic, molecular, and charged-particle sources and detectors
 % 05.30.Jp, Boson Systems
 % 42.79.Gn, Optical waveguides and couplers (for fiber waveguides and waveguides in integrated optics, see 42.81.Qb and 42.82.Et, respectively)
 % 42.65.Jx, Beam trapping, self-focusing and defocusing; self-phase modulation
 % 42.81.Qb, Optical waveguides and couplers (for fiber waveguides and waveguides in integrated optics, see 42.81.Qb and 42.82.Et, respectively)
 % 43.20.Mv ???
 % 03.75.Lm Tunneling, Josephson effect, Bose-Einstein condensates in periodic potentials, solitons, vortices and topological excitations
 % 03.75.Hh Static properties of condensates; thermodynamical, statistical and structural properties
 % 03.75.Fi, 74.50.+r, 05.30.Jp, 32.80.Pj 
 % 03.75.Kk, 03.75.Lm, 03.65.Sq
 % 42.79.Fm Reflectors, beam splitters, and deflectors
 % 42.82.Gw Other integrated-optical elements and systems
 % 42.65.Dr Stimulated Raman scattering; CARS
 % 42.65.Es Stimulated Brillouin and Rayleigh scattering
 % 42.82.Cr Fabrication techniques; lithography, pattern transfer

\section{INTRODUCTION}

Since the creation of the first Bose-Einstein condensate (BEC)~\cite{Anderson1995Ob,Davis1995Bo}, cold atom experiments have provided a powerful
platform for the study of macroscopic quantum states~\cite{Raghavan1999Co,Huang2006Cr}, emulation of solid state
physics~\cite{Lewenstein2007Ul,Bloch2008Ma}, and insight into many-body quantum phenomena. The utility of BECs stems from the experimental freedom to
control many system parameters, including the interaction strength, degrees of freedom, size, and shape of the BEC.

One enticing proposal is to use BECs for interferometry~\cite{Wang2005At,Schumm2005Ma,Shin2004At}. BEC interferometers would have many advantages over
their optical counterparts. Trapped-atom interferometers can be sensitive to changes in mass, charge, magnetic moment and polarisability. As with
optical interferometry, a BEC interferometer involves the spatial splitting, followed by the generation of a relative phase difference between split
components and then coherent recombination of the quantum state. Performing these operations in a way that is relatively insensitive to small errors
in the implementation is a necessary precondition for a reliable interferometer.

The conventional two-well splitting of a BEC is highly sensitive to atom-atom interactions, where phase diffusion and Josephson oscillations lead to a
loss of phase resolution~\cite{Castin1997Re,Jo2007Ph,Li2007Ex}.  Beam splitting via laser-induced adiabatic passage has the advantage of being robust
to changes in experimental parameters.  As first proposed by Marte \emph{et al}~\cite{Marte1991Co} a variation of the efficient and robust three-state
adiabatic process known as stimulated Raman adiabatic passage~(STIRAP) \cite{Oreg1984Ad,Kuklinski1989Ad,Gaubatz1990Po,Bergmann1998Co}, called
fractional STIRAP (f-STIRAP) \cite{Lawall1994De,Weitz1994At}, can be used to generate any preselected coherent superposition of two atomic states,
$\ket{1}$ and $\ket{3}$, via an intermediate excited state, $\ket{2}$. Electromagnetic pulses are used to couple states $\ket{1}$ to $\ket{2}$ and
$\ket{2}$ to $\ket{3}$, characterised by coupling parameters $\Omega_{12}$ and $\Omega_{23}$. As in STIRAP, $\Omega_{23}$ precedes $\Omega_{12}$ but
unlike STIRAP where $\Omega_{23}$ vanishes first, here the two pulses vanish simultaneously while maintaining a constant ratio of amplitudes. The
ratio of probability amplitudes of the resulting coherent superposition of states $\ket{1}$ and $\ket{3}$ is proportional to the ratio
$\Omega_{23}/\Omega_{12}$. Hence stopping at $\Omega_{23}/\Omega_{12}=1$ can create an atomic beam splitter, as demonstrated experimentally by Weitz
\emph{et al}~\cite{Weitz1994Ata}.

Here we propose an alternative method for the spatial splitting and coherent recombination of a BEC based on the ideas underpinning f-STIRAP. In
three-well atomic~\cite{Eckert2004Th} and electronic quantum dot~\cite{Greentree2004Co} systems, the coherent spatial transport of single particle
quantum states is known as coherent tunnelling adiabatic passage~(CTAP). Recent work has shown that this principle can be extended to interacting
many-body quantum systems such as BECs, both in the quantum~\cite{Bradly2012Co} and semi-classical mean-field
limits~\cite{Nesterenko2009ST,Liu2007Jo,Wang2006La,Itin2007In,Graefe2006Me,Rab2008Sp,OSullivan2010Us}.

\begin{figure}[tb!]
  \centering
  \includegraphics[width=\columnwidth]{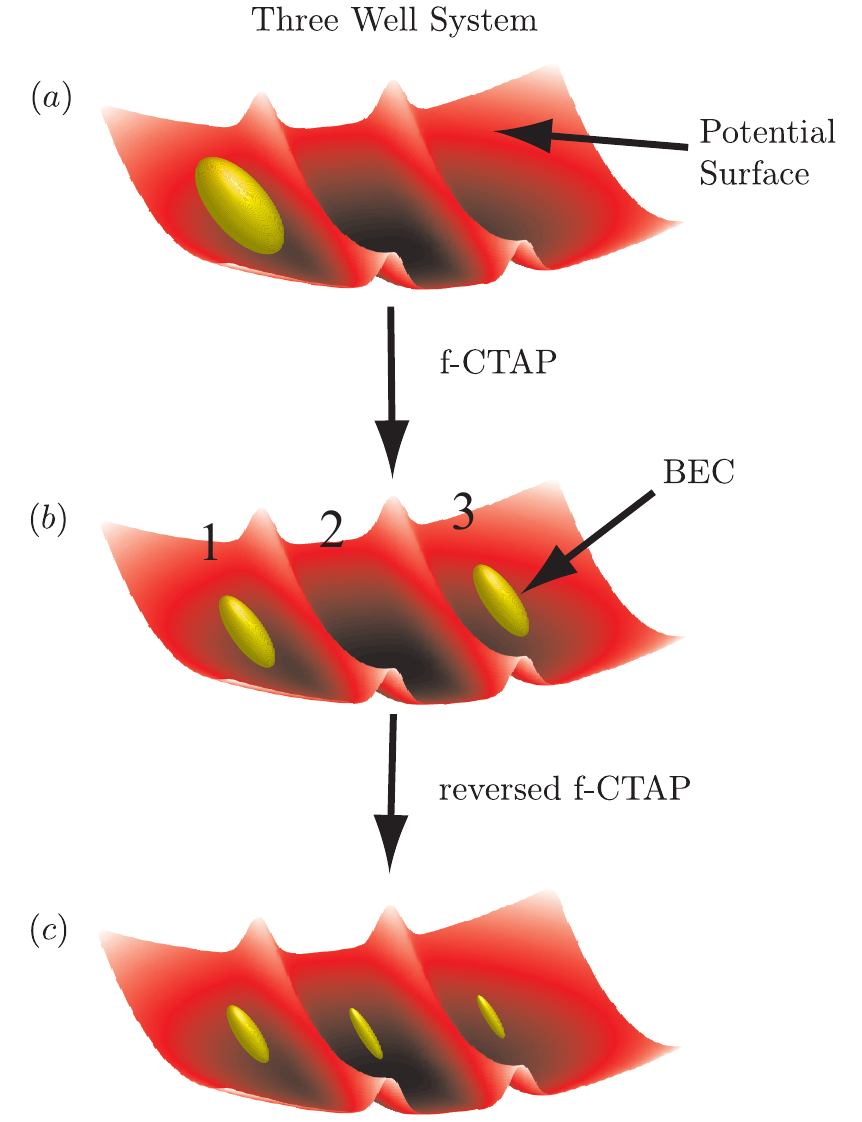}
  \caption{(Color online) Schematic representation of three-well system: at (a) $t=0$, at (b) $t_p/2 < t <t_p/2 + \tau$ where $\tau$ is the hold time
    required for phase accumulation, and at (c) $t=t_p + \tau$. The system consists of two parallel, repulsive Gaussian barriers embedded in an
    ambient harmonic trap, dividing the system into three wells. At $t=0$ - (a) - the BEC resides in well . At $t=t_p/2$ - (b) - the BEC is split into
    two equal components residing in wells 1 and 3. To perform the interferometric measurement, the tunnelling rates are then kept constant for some
    hold time, $\tau$, during which a relative phase difference may be accumulated. At $t=t_p+\tau$ - (c) - the two BECs are recombined which leads to
    a phase-dependent population in well 1.}
  \label{System}
  \vspace{-0.0cm} \end{figure}

Analogously to f-STIRAP, fractional CTAP~(f-CTAP) allows for the creation of a coherent spatial superposition of eigenstates of wells 1 and 3. For
photons, f-CTAP has been demonstrated experimentally in three-channel optical waveguides~\cite{Dreisow2009Po,Menchon-Enrich2012Ad}. Consider the
initial state with the BEC confined to well 1, schematically shown in Fig.~\ref{System}(a). The atomic population of the BEC is transported from well
1 to 3 through adiabatic changes to the tunnelling rates between the wells, and is halted once the BEC is equally split [see Fig.~\ref{System}(b)]. At
this point, one component of the BEC can be made to interact with some system of interest for time, $\tau$. In the case where a phase difference
accumulates between the two states of the superposition, this phase can be determined via an interferometric measurement, as demonstrated in two-well
systems~\cite{Hofferberth2006Ra,Shin2004At} by releasing the BEC from the three-well system. In this work, f-CTAP is proposed to be also used in the
recombination stage of the interferometer as reversing the f-CTAP protocol results in a phase-dependent population of the three wells~[see
Fig.~\ref{System}(c)]. Interferometric f-CTAP needs to be performed on a timescale that is much longer than non-adiabatic methods. However, the
tunnelling interaction between the split BECs maintains mutual coherence throughout the splitting and recombination processes. As with CTAP, the
f-CTAP protocol has the advantage of being robust to variations in trap parameters and pulse time.

To explore the dynamics of our three-well system, we employ the three-mode Gross-Pitaevskii equation
(GPE)~\cite{Graefe2006Me,Wang2006La,Liu2007Jo,Zhang2001In,Franzosi2003Ch,Ottaviani2010Ad}. In the noninteracting limit we show analytically that a BEC
initially residing in well 1 can be split between wells 1 and 3, and recombined to give a phase dependent density in well 1~[see Fig.~\ref{System}].
To understand the role of interactions we use a nonlinear three-mode treatment.  The presence of interactions gives rise to a window where canonical
CTAP can occur and also a regime where more sensitive interferometric sensing can be observed. By solving the corresponding classical equations of
motion, we identify the nonlinear eigenstates of the system, identifying the bifurcation point where extra nonlinear states appear, and investigate
the stability of the CTAP state.

The nonlinear dynamics that are obtained from the interaction CTAP model have been mapped directly onto a corresponding three-dimensional
GPE~\cite{Rab2008Sp} showing that the adiabatic transport of a BEC containing 2000 $\mathrm{^7Li}$ atoms can be achieved over $20\si{\micro\meter}$
within an ambient harmonic trap of $\omega = 2\pi\times40\si{\hertz}$.

\section{Noninteracting Modal Approximation}

Consider a three-well system. For sufficiently large wells, the system
is accurately described by the modal
approximation~\cite{Smerzi1997Qu,Graefe2006Me,Wang2006La,Oreg1984Ad,Liu2007Jo,Ottaviani2010Ad}. Each well is approximated by a single
mode, $\Psi_i$, giving:
\begin{equation}
  \Psi=\sum_{j=1}^3 \psi_j(t)\Phi_j(x),
\end{equation}
where the amplitude of each mode $j$ is expressed as $\psi_j=\sqrt{N_j}e^{i\phi_j}$, with $N_j$ and $\phi_j$ being the
modal occupation and phase respectively. The system is normalised such that $\sum_{j=1}^3 N_j(t)=1$. In this basis, the
Hamiltonian of the three-well system is:
\begin{align}
  H=\hbar\Omega\left(
    \begin{array}{ccccc}
      U_1 & -\Omega_{12} & 0 \\
      -\Omega_{12} & U_2 & -\Omega_{23}\\
      0 & -\Omega_{23} & U_3 \\
    \end{array}
  \right), \label{CTAP_H}
\end{align}
where $\Omega_{ij}$ is the dimensionless tunnelling rate between modes $i$ and $j$, determined by the wavefunction overlap. The
dimensionless on-site interaction energy per particle is $U_j=E_j^0+g N_j$.
$\hbar{}\Omega{}E_j^0$ is the groundstate energy of well
$j$, with $\Omega$ being the maximum tunnelling rate that sets the
characteristic energy scale of the system,  and $g$ the dimensionless parameter describing
the nonlinear atomic interactions.

We first describe the f-CTAP protocol for BEC splitting and recombination in the noninteracting regime, $g = 0$.

\subsection{BEC Splitting with Fractional CTAP}

\begin{figure}[tb!]
  \centering
  \includegraphics[width=\columnwidth]{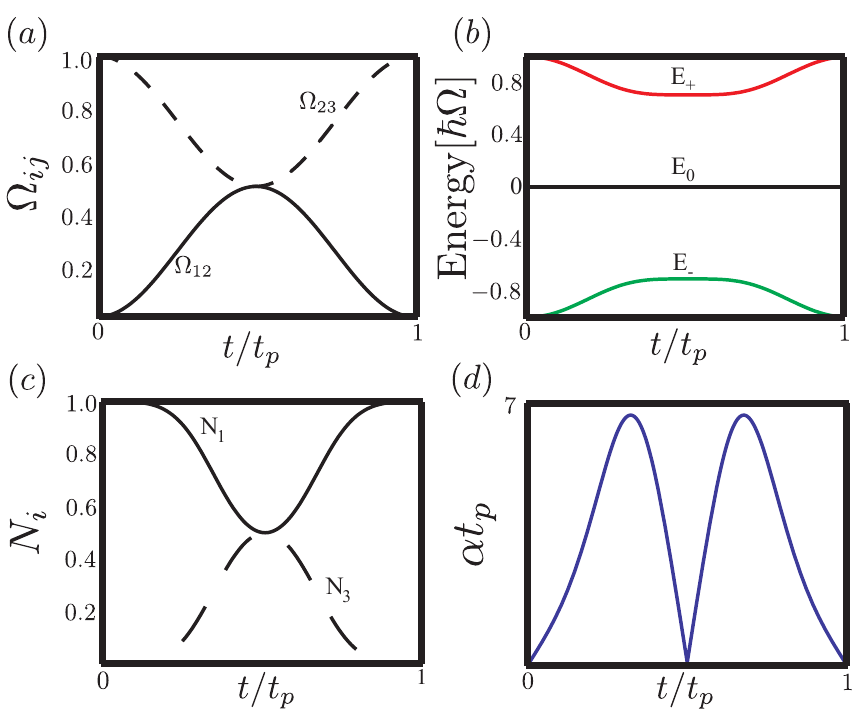}
  \caption{(Color online) Ideal three-well system with $\Delta=0$ and $g=0$: (a) Proposed pulsing scheme: Eqs.~\eqref{K1_3} and \eqref{K2_3}. (b)
    Energies of the eigenmodes: $D_0$ and $D_{\pm}$. (c) Evolution of the occupation of the three wells: $N_1$ (solid curve) and $N_3$ (dashed
    curve). (d) Adiabaticity parameter, $\alpha t_p$. In these figures we have assumed that $\tau=0$.}
  \vspace{-0.0cm}
  \label{fig:linear}
\end{figure}

In the absence of nonlinear atomic interactions ($g = 0$) and for $E_1^0=E_3^0=0$, $E_2^0=\Delta$, the eigenstates of the Hamiltonian,
Eq.~\eqref{CTAP_H}, are:
\begin{align}
  D_0&=\frac{\Omega_{23}\Phi_1}{\sqrt{\Omega_{23}^2+\Omega_{12}^2}} -\frac{\Omega_{12} \Phi_3}{\sqrt{\Omega_{23}^2+\Omega_{12}^2}}, \\
  D_{\pm}&=\frac{1}{\sqrt{\gamma(\gamma \pm \Delta) }}\left[\sqrt{2} \Omega_{12}\Phi_1-\frac{\Delta\pm\gamma}{\sqrt{2}} \Phi_2 +
    \sqrt{2}\Omega_{23}\Phi_3\right],
\end{align}
where $\gamma = \sqrt{\Delta^2 + 4(\Omega_{12}^2+\Omega_{23}^2)}$. The $D_0$ eigenstate has no population in the centre well for all $\Delta$ and in
the limit of $\Omega_{12}/\Omega_{23} \ll 1$ the atomic population is confined entirely to well 1.

Consider the scenario where the BEC initially resides in well 1 ($\Omega_{12}/\Omega_{23} = 0$). Adiabatic transport through the $D_0$ state such that
the BEC is equally split between wells 1 and 3, at $t=t_p/2$ requires a smooth transformation of $\Omega_{12} = 0$ to $\Omega_{12}=\Omega_{23} =
1/2$. A pulsing sequence, for $\tau=0$ that meets this criterion is:
\begin{align}
  \Omega_{12}(t) &= \frac{1}{2}\sin^2\left[\pi t/t_p\right],  \label{K1_3}\\
  \Omega_{23}(t)&=\frac{1}{2}+\frac{1}{2}\cos^2\left[\pi t/t_p\right], \label{K2_3}
\end{align}
as shown in Fig.~\ref{fig:linear}(a). The mode energies for this pulse sequence are shown in Fig.~\ref{fig:linear}(b) with the population of the three
wells shown in Fig.~\ref{fig:linear}(c) (where adiabatic pulsing has been assumed). At $t_p/2$, $N_1=N_3=0.5$ which is the condition we require for
sensing.

The robustness of the splitting protocol relies on adiabatic changes to the system tunnelling. For noninteracting BECs, the adiabaticity is quantified
by the adiabaticity parameter:
\begin{equation}
  \alpha =
  \max_{i=\pm}\frac{\left|\langle{}D_0;t|\frac{\partial}{\partial t}H(t)|D_i;t\rangle\right|}{\left|E_0(t)-E_i(t)\right|^2},
  \label{equ:adiabaticcondition}
\end{equation}
which is shown in Fig.~\ref{fig:linear}(d) through the protocol. Adiabatic transfer requires $t_p$ to be chosen such that $\alpha \ll 1$ over the
entirety of the protocol.

\subsection{Phase Interferometry with Fractional CTAP}
\label{sec:nonintphase}

Once the BEC is split, interaction with a target system can induce a relative phase difference between the two components. Since
$\Omega_{12}=\Omega_{23}=1/2$ at $t=t_p/2$, phase information is shared between the BECs in wells 1 and 3. This enables the mutual coherence between
the two BECs during splitting to be maintained, inhibiting phase diffusion.  To enable the desired phase accumulation, tunnelling between wells 1 and
3 needs to be suppressed. This can be achieved via a variety of means, such as (i) applying a blocking laser field at the middle point of well 2 or
(ii) reducing $\Omega_{12}$ and $\Omega_{23}$ adiabatically to zero. After the decoupling of the BECs, phase diffusion will occur. The BECs can be
recoupled some time later by, for example, (i) switching off the blocking laser field or (ii) adiabatically increasing $\Omega_{12}$ and $\Omega_{23}$
to $1/2$. We assume that the process of decoupling and recoupling occurs over timescale $\tau$. Considering the case where the two BECs accumulate a
phase difference $\phi$ over $\tau$, the state of the system is then:
\begin{equation}
  \Psi_S(t_p/2+\tau,\phi) = \frac{1}{\sqrt{2}}\left(\Phi_1 - e^{i\phi}\Phi_3\right).
  \label{Psi_S_three_well}
\end{equation}
The acquired phase difference during $\tau$ moves the system out of the $D_0$ state, and is now in a superposition of all three eigenstates. Hence,
the resulting change in eigenstate populations for a given $\phi$ is:
\begin{align}
  N(D_+) &= \sin^2(\phi/2) \sin^2 \theta \label{ND_p},\\
  N(D_-) &=\sin^2(\phi/2) \cos^2 \theta \label{ND_m},\\
  N(D_0)&=\cos^2 (\phi/2) \label{ND_0},
\end{align}
where
\begin{equation}
  \theta =\frac{1}{2}\arctan\left[\sqrt{2}/\Delta\right].
\end{equation}
Adiabatic evolution of the system through to $t=t_p+\tau$ returns the barriers back to their initial configuration, where $\Omega_{12} = 0$ and
$\Omega_{23} = 1$. At the end of the splitting and phase accumulation ($t=t_p+\tau$), the $D_0$ eigenstate is simply the state in well 1 with the
final population given by:
\begin{equation}
  N_1(t_p+\tau)=N(D_0)=\cos^2 (\phi/2).
  \label{eq:1}
\end{equation}
Thus, the phase difference accumulated during $\tau$ manifests as a population imbalance at the end of the protocol. Density measurements on the final
state of the BEC then allow determination of the phase, $\phi$. This is demonstrated by full numerical integration of the three-mode GPE in
Section~\ref{sec:interferometry} where the $g=0$ lines in Fig.~\ref{fig:tp100-3000} confirms the phase response predicted by Eq.~\eqref{eq:1}.

Noise in the tunnelling throughout the transport can be suppressed by considering an experimental setup where a single laser is split proportionally
between two barriers to mediate the tunnelling interaction. Hence, noise in $\Omega_{12}$ and $\Omega_{23}$ due to fluctuations in the intensity of the
source laser will be coupled. As shown in Eckert \textit{et al} \cite{Eckert2004Th}, adiabatic transport is still achievable in the presence of such
noise. 

Any small asymmetry in the final tunnelling terms of the Hamiltonian will result in a loss of sensitivity proportional to the population difference
between wells $\ket{1}$ and $\ket{3}$, as such does not affect our analysis.

\section{Nonlinear Three-Mode Approximation}
\label{sec:interactingH}

Controllable nonlinearities are one of the defining features of BEC dynamics as compared to analogous photonic systems. Accordingly we now consider
the effect of a non-zero $g$ on the f-CTAP interferometric protocol. Inter-particle interaction in BECs gives rise to a raft of interesting phenomena,
for example quantum phase transitions~\cite{Bloch2008Ma}, solitons~\cite{Burger1999Da,Khaykovich2002Fo,Strecker2002Fo,Cornish2006Fo}, and entanglement
generation~\cite{Romero-Isart2007Tr}. However, in the context of BEC interferometry, these interactions are in general problematic, as they induce phase diffusion. These
interactions manifest as a nonlinearity in the mean-field description, leading to highly non-trivial dynamics.  We analyse the behavior of the system
by mapping the nonlinear GPE to a classical Hamiltonian. The nature of the stationary states in the classical model determines the behaviour of the
protocol in the adiabatic limit. This analysis is complimented by numerical simulation. Simulation of the protocol confirms the conclusions of the
classical analysis in the large $t_p$ limit, and reveals several interesting phenomena for finite $t_p$.  Interestingly, we find that the presence of
nonlinearities can lead to an \emph{enhanced} phase sensitivity.

\subsection{Mapping to Classical Variables}

The presence of a nonlinearity in the GPE makes analysis of the dynamics significantly more difficult. Notably, the superposition principle is no
longer valid, making an eigenstate decomposition impossible. In general there will be more eigenstates than the Hilbert space dimension. The
system can be approached though concepts borrowed from classical mechanics, as there is a mapping of our system to a classical Hamiltonian. From this
perspective, we study the stationary points in the classical phase space, that correspond to the eigenstates of the original system.

Any quantum system with Hilbert space dimension $N$ can be mapped onto a classical system with $2N$ degrees of freedom, namely, the $N$ phases and $N$
amplitudes of the state vector. Symmetry under global phase shifts leads to a conservation of probability amplitude as an integral of motion, and if
the Hamiltonian is time-independent, the total energy is also an integral of motion.  This implies that two-mode BECs have non-chaotic dynamics. In
the three-well case, integrable dynamics are no longer guaranteed. The presence of chaotic dynamics has implications for the splitting and
recombination elements of the interferometer as adiabatic transport through a chaotic region of phase space is not possible. Linear quantum systems
map onto $N$-dimensional harmonic oscillators, and are necessarily integrable. Introduction of nonlinear terms leads to more complicated classical
dynamics, which in some cases can be chaotic. This is indeed the case for the three-well BEC~\cite{Nemoto2001Qu,Thommen2003Cl}.

We take the classical degrees of freedom to be the amplitude squared and phase of the BEC in each well: $N_i = \left|\psi_i\right|^2$, $\phi_i =
\arg{(\psi_i)}$. The number of degrees of freedom can be reduced by two using conservation of probability and global phase symmetry:
\begin{align}
  \begin{split}
    N_2 & =  1 -N_1 - N_3, \\
    \phi_{12} & =  \phi_1 - \phi_2,  \\
    \phi_{32} & = \phi_3 - \phi_2.
  \end{split}
  \label{equ:cdof}
\end{align}
The classical Hamiltonian is then:
\begin{equation}
  \begin{aligned}
    H/\hbar\Omega&=\Delta(1 - N_1 - N_3) \\
    & \quad  + \frac{g}{2}[N_1^2 + N_3^2 + (1-N_1-N_3)^2]\\
    & \quad   -  2\Omega_{{12}}\sqrt{N_1(1- N_1 -   N_3)}\cos{\phi_{12}}\\
    & \quad - 2\Omega_{23}\sqrt{N_3(1 - N_1 - N_3)}\cos{\phi_{32}},
    \label{equ:cham}
  \end{aligned}
\end{equation}
and the equations of motion are:
\begin{align}
  \begin{split}
    \dot{\phi_{12}}/\hbar\Omega&=-\frac{\Omega_{12}(2N_1 + N_3 - 1)\cos{\phi_{12}}}{\sqrt{N_1(1-N_1-N_3)}}  -\Delta\\
    &\quad- \frac{\Omega_{23}N_3\cos{\phi_{32}}}{\sqrt{N_3(1-N_1-N_3)}} + g(2N_1 + N_3 - 1),
  \end{split}\\
  \dot{N_1}/\hbar\Omega &= 2\Omega_{12}\sqrt{N_1(1- N_1-N_3)}\sin{\phi_{12}},\\
  \begin{split}
    \dot{\phi_{32}}/\hbar\Omega &= -\frac{\Omega_{23}(N_1 + 2N_3 - 1)\cos{\phi_{32}}}{\sqrt{N_3(1-N_1-N_2)}}  -\Delta\\
    &\quad - \frac{\Omega_{12}N_1\cos{\phi_{12}}}{\sqrt{N_1(1-N_1-N_3)}} + g(N_1 + 2N_3 - 1),
  \end{split}\\
  \dot{N_3}/\hbar\Omega &= 2\Omega_{23}\sqrt{N_3(1- N_1-N_3)}\sin{\phi_{32}}.
  \label{equ:ceom}
\end{align}

\subsubsection{Stationary States}
\label{sec:nonl-eigenst-1}

CTAP, by definition, requires evolution along particular eigenstates. In Section~\ref{sec:nonintphase} we were able understand sensing by the
accumulated phase shifting the BEC into a superposition of all three eigenstates. However, the nonlinearities in the modal BEC approach invalidate the
superposition principle. Nevertheless studying the states of the system still yields useful insight into the structure of the resulting phase space.

\begin{figure}[tb!]  \centering
  \includegraphics[width=\columnwidth]{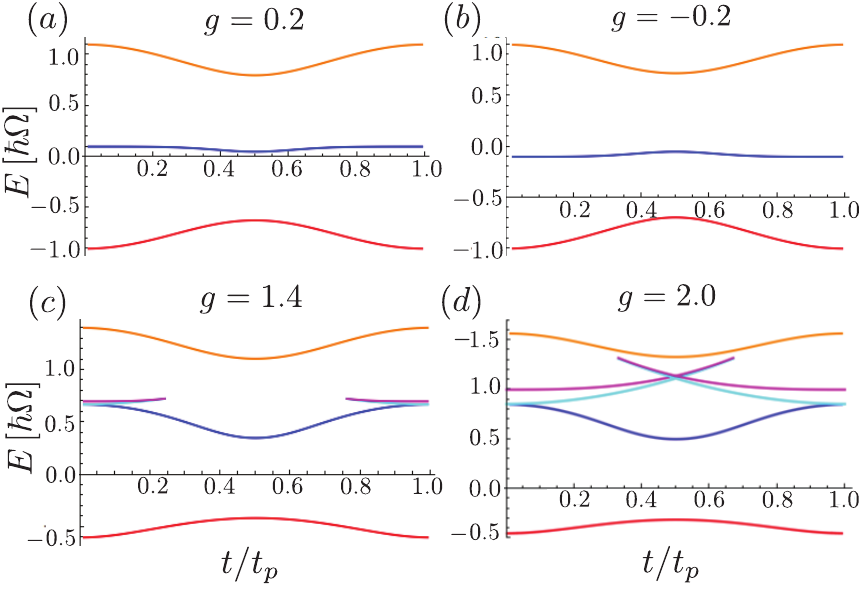}
  \caption{(Color online) Eigenenergies for a range of interaction strengths with $\Delta = 0.1$ and $t_p\rightarrow\infty$. As the interaction
    strength is increased, new eigenstates appear. When $|g|\approx{}1.0$, the $D^{\prime}_0$ state at $t=0$ is no longer $\psi_1$, precluding the
    possibility of a 50/50 split. }
  \vspace{0.0cm}
  \label{fig:eigs}
\end{figure}

The eigenvalues are given by the stationary solutions to Eq.~\eqref{equ:ceom}. To distinguish from the $g=0$ case we denote these stationary states by
$D^{\prime}$, with the $D^{\prime}_0$ coinciding with the $D_0$ split state at $t=t_p/2$:
\begin{equation}D^{\prime}_0(t_p/2)= \frac{1}{\sqrt{2}}\left(\Phi_1-\Phi_3\right)= \Psi_{50/50}.
  \label{50/50}
\end{equation}
As mentioned above for the interacting case, the number of eigenvalues is no longer limited to three~(see Ref.~\cite{Dagosta2002St} for an in-depth
discussion).  Fig.~\ref{fig:eigs} shows the eigenvalues for various values of $g$, using the pulsing sequence defined by
Eqs.~\eqref{K1_3}~and~\eqref{K2_3}.  At higher interaction strengths, new eigenstates appear. The appearance of these new states marks a bifurcation
near the $t=\left\{0,t_p\right\}$ limits, which disconnects the $D^{\prime}_0$ state from the fully occupied states at the end points. Once these
eigenstates appear at $g = g_c = \Delta/2 \pm \sqrt{1 + \Delta^2/4}$, 50/50 splitting is precluded, even in principle.

The appearance of these additional nonlinear eigenstates is a ubiquitous feature of nonlinear systems. However, their effect on the dynamics depends
on the pulse sequence that is employed. In particular Graefe \emph{et al} ~\cite{Graefe2006Me} showed that the extra eigenstates are a permanent
feature of the eigenspectrum for all $g\Delta\le 0$ and $|g| > |\Delta|$ using a Gaussian pulsing protocol, precluding adiabatic transport. Conversely
$\Omega_{23}\ne0$ at the start of the protocol, as used here, has the effect of suppressing the emergence of these extra nonlinear eigenstates for low
values of $|g|$, as seen in the window of robust evolution in Fig.~\ref{fig:fidelity}. This robustness is further considered in the stability analysis
below.

\begin{figure}[tb!]  \centering
  \includegraphics[width=\columnwidth]{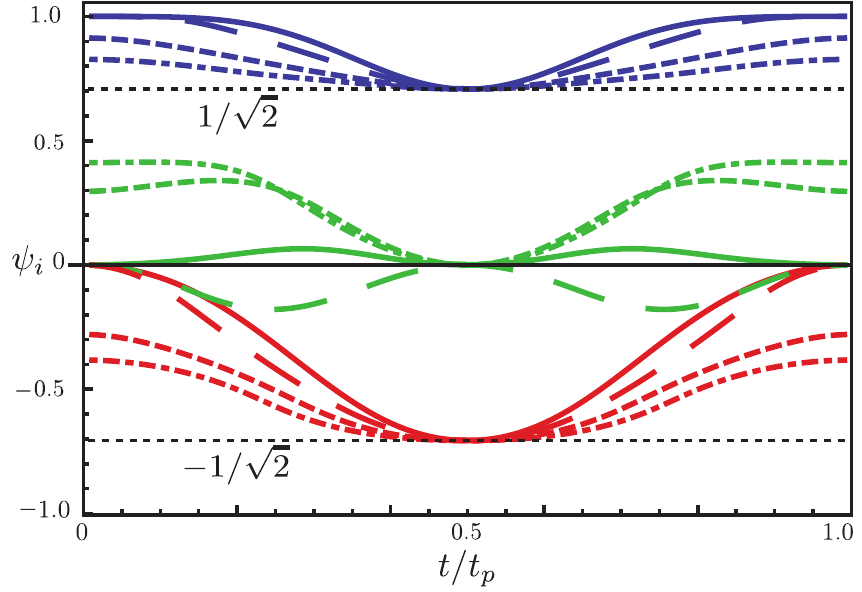}
  \caption{(Color online) Wavefunction components for the $D^{\prime}_0$ state with non-zero $g$, $\Delta=0.1$ and $t_p\rightarrow\infty$.  Upper(blue):
    $\psi_1$; middle(green): $\psi_2$; lower(red): $\psi_3$. Solid: $g=0.2$; long dash: $g=-0.6$; short dash: $g=1.4$; dot-dash: $g=2.0$.  Interactions lead to
    occupation of the middle well during transport, as can be seen for $g=0.7$ and $g=1.0$. Note: $\psi_i$ is real.}
  \vspace{-0.0cm}
  \label{fig:eigv}
\end{figure}
From Eq.~\eqref{equ:ceom} it is possible to calculate the wavefunction amplitudes in the three wells as a function of the tunnelling rates and
nonlinear interaction. Fig.~\ref{fig:eigv} shows that for $g=0$, $\psi_2 = 0$ for all $t$. However, upon the introduction of interactions $\psi_2$
will in general be non-zero. After the appearance of the self-trapped mode, the 50/50 split state at $t/t_p = 0.5$ is no longer adiabatically
connected to the fully-occupied well 1 state at $t/t_p = 0$.

\subsubsection{Stability}
\label{sec:stability}
We now consider the stability of the $D^{\prime}_0$ state in the regime of $|g| < |g_c|$. A stationary point is unstable if small perturbations lead
to large deviations from the point. These deviations make adiabatic transport of the system impossible, as keeping the system near $D^{\prime}_0$
requires that $t_p\rightarrow\infty$.

When $|g| < |g_c|$, there are just the three eigenstates. However, these differ from the noninteracting case, i.e. the $D^{\prime}_0$ state's
eigenenergy is no longer zero and, in general, $\psi_2(t)\ne 0$, as demonstrated in Fig.~\ref{fig:eigv}.  The dynamics in the neighbourhood of
stationary states can be studied by making a linear expansion of the equations of motion. Solutions are classified by eigenvalues of the Jacobian at
the stationary point:
\begin{equation} J_{ij} = \frac{\partial{}\dot{x}_i}{\partial{}x_j},
  \label{}
\end{equation}
where the $x_i$ are the system variables. If the eigenvalues are purely imaginary, then the solutions are oscillatory around the stationary point. An
eigenvalue with a positive real part implies the presence of a hyperbolic orbit around the eigenstate, which is unstable.

\begin{figure}[tb!]
  \centering
  \includegraphics[width=\columnwidth]{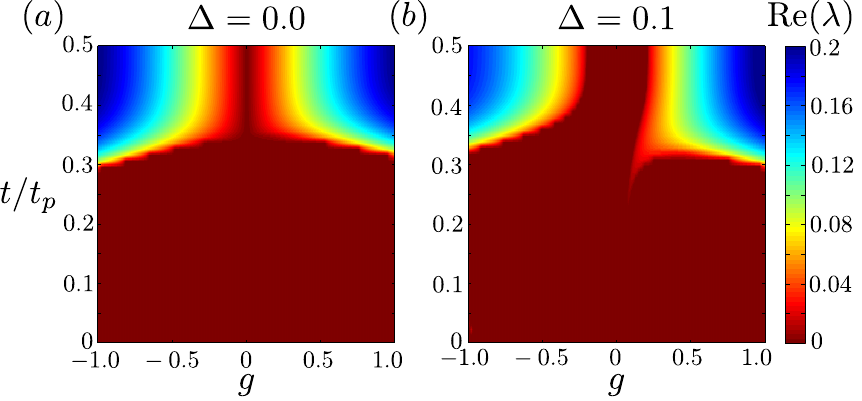}
  \caption{(Color online) Real part of the eigenvalue of the Jacobian at the $D^{\prime}_0$ state. (a) $\Delta = 0.0$. (b) $\Delta = 0.1$.  Transport through regions
    with $\textrm{Re}(\lambda) = 0 $ allows complete fidelity for the protocol. For $\Delta = 0$, the $D^{\prime}_0$ state is unstable for all $g \ne 0$,
    albeit with small $\textrm{Re}(\lambda)$. }
  \vspace{-0.0cm}
  \label{fig:lypdark}
\end{figure}

Fig.~\ref{fig:lypdark} shows how the stability of the $D^{\prime}_0$ state changes along the pulse sequence for $-1 < g < 1$, for $\Delta=0$
\ref{fig:lypdark}(a) and $\Delta=0.1$ \ref{fig:lypdark}(b). For $\Delta=0$ the $D^{\prime}_0$ state is strictly only stable for $g=0$ throughout the
pulse sequence. However, the introduction of $\Delta$ stabilises the $D^{\prime}_0$ state throughout the transport protocol. Specifically, for
$\Delta=0.1$, transport is stable in the range $|g|\alt0.2 $ and, as can be seen in Fig.~\ref{fig:eigv}, complete fidelity for transport to a split
state is possible for extended pulse times. At higher $|g|$, transporting through regions of instability leads to a loss of fidelity. For an
intermediate interaction strength, this instability is small, and it is possible to find a balance between the adiabaticity requirement and the slow
divergence of the $D^{\prime}_0$ state which still allows high fidelity for the splitting protocol. Hence we have identified $\Delta=0.1$ as
advantageous to stable transport. We adopt this parameter in the following analysis.

\subsection{Nonlinear Fidelity of Splitting}
We now explore the robustness of the f-CTAP splitting protocol by direct numerical evolution of the nonlinear Hamiltonian. Specifically we solve for
$\psi_j$ in the presence of nonlinear interactions and finite pulsing times, via:
\begin{equation}
  \frac{i}{\Omega}\frac{\partial}{\partial
    t}\Psi=H\Psi.
  \label{eq:numerical}
\end{equation}
We are interested in the transportation of the BEC to the split state. We therefore define the fidelity as the overlap between the $\Psi_{50/50}$
state and the transported state, $\Psi(t_p/2)$, as determined from Eq.~\eqref{eq:numerical}:
\begin{equation}
  \epsilon = \left|\langle\Psi_{50/50}|\Psi(t_p/2)\rangle\right|^2.\label{eq:epsilon}
\end{equation}

\begin{figure}[tb!]
  \centering
  \includegraphics[width=0.5\textwidth]{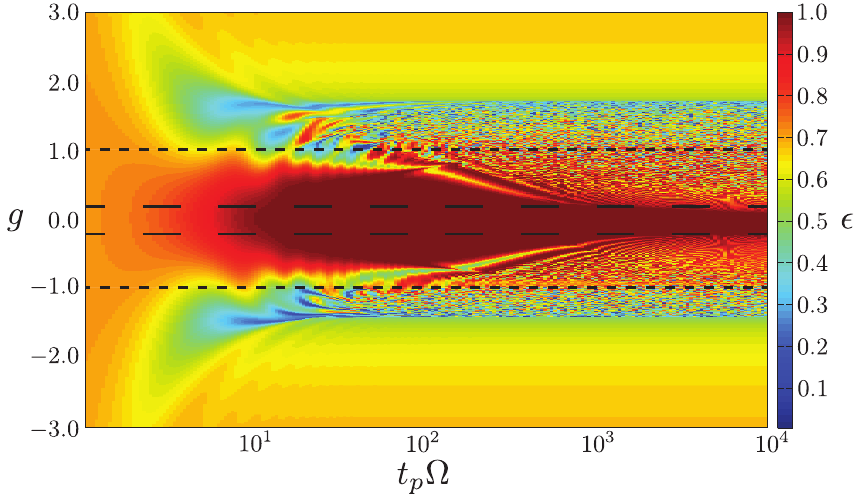}
  \caption{(Color online) Fidelity, $\epsilon$, of split as defined in Eq.~\eqref{eq:epsilon} as a function of pulse time, $t_p$, and interaction strength, $g$, at
    $t=t_p/2$ with $\Delta=0.1$. Regions of good fidelity, $\epsilon\approx1$, are denoted by dark red. The short dashed lines denote $g=g_c$ where
    the bifurcation appears and the long dashed lines denote $|g|=0.2$.}
  \vspace{-0.0cm}
  \label{fig:fidelity}
\end{figure}

Fig.~\ref{fig:fidelity} shows the fidelity of BEC splitting in the presence of interactions. Qualitatively, for small values of $|g|$, full adiabatic
transport is still possible for $t_p/2>10\Omega^{-1}$. However, larger interaction strengths lead to oscillations and complete loss of fidelity in
the splitting procedure.

The regions of efficient splitting in Fig.~\ref{fig:fidelity} can be explained in terms of the stability of the $D^{\prime}_0$ state
(Fig.~\ref{fig:lypdark}). For the pulsing scheme given by Eqs.~\eqref{K1_3} and \eqref{K2_3}, stable transport to the split state is achieved for $|g|
< 0.2$ in the adiabatic limit (long dashed lines in Fig.~\ref{fig:fidelity}). This region is characterised by its close-to-linear behaviour, where an
increase in total pulse time leads to a corresponding increase in fidelity. The small instability that exists for positive $g$ slightly reduces the
fidelity of transport for long pulse times. The absence of extra nonlinear states (Fig.~\ref{fig:eigs}) and stability of the $D^{\prime}_0$ state in
this range (Fig.~\ref{fig:lypdark}) means that the fidelity of transport obeys the linear adiabatic law up to very long pulse times.

For $|g|>0.2$, transport for the $D^{\prime}_0$ state becomes unstable (Fig.~\ref{fig:lypdark}), with the growth of this instability increasing with
$|g|$. To achieve good transport in this regime, the optimal pulse time must be found. This comes from competition between the adiabaticity of the
transport protocol and the instability timescale of the $D^{\prime}_0$ state, with shorter $t_p/2$ being favoured for stronger interactions. At $g =
g_c$, the appearance of extra nonlinear states near the $D^{\prime}_0$ state, which only extend partially along the pulsing scheme, as shown in
Fig.~\ref{fig:eigs}, prevent stable splitting for any $t_p/2$. Once these extra states extend to $t=t_p/2$, stability in transport is
restored. However, the final state does not overlap the $\Psi_{50/50}$ state.

The evolution shown in Fig.~\ref{fig:fidelity} is the first stage of this interferometric process and reversal of the protocol, for recombination,
will naturally lead to high fidelity transport analogous to full CTAP transport. The full sensing protocol is the subject of the next section.

\subsection{Phase Interferometry in the Presence of Interactions}
\begin{figure*}[tb!]
  \begin{center}
    \includegraphics[width=1.00\textwidth]{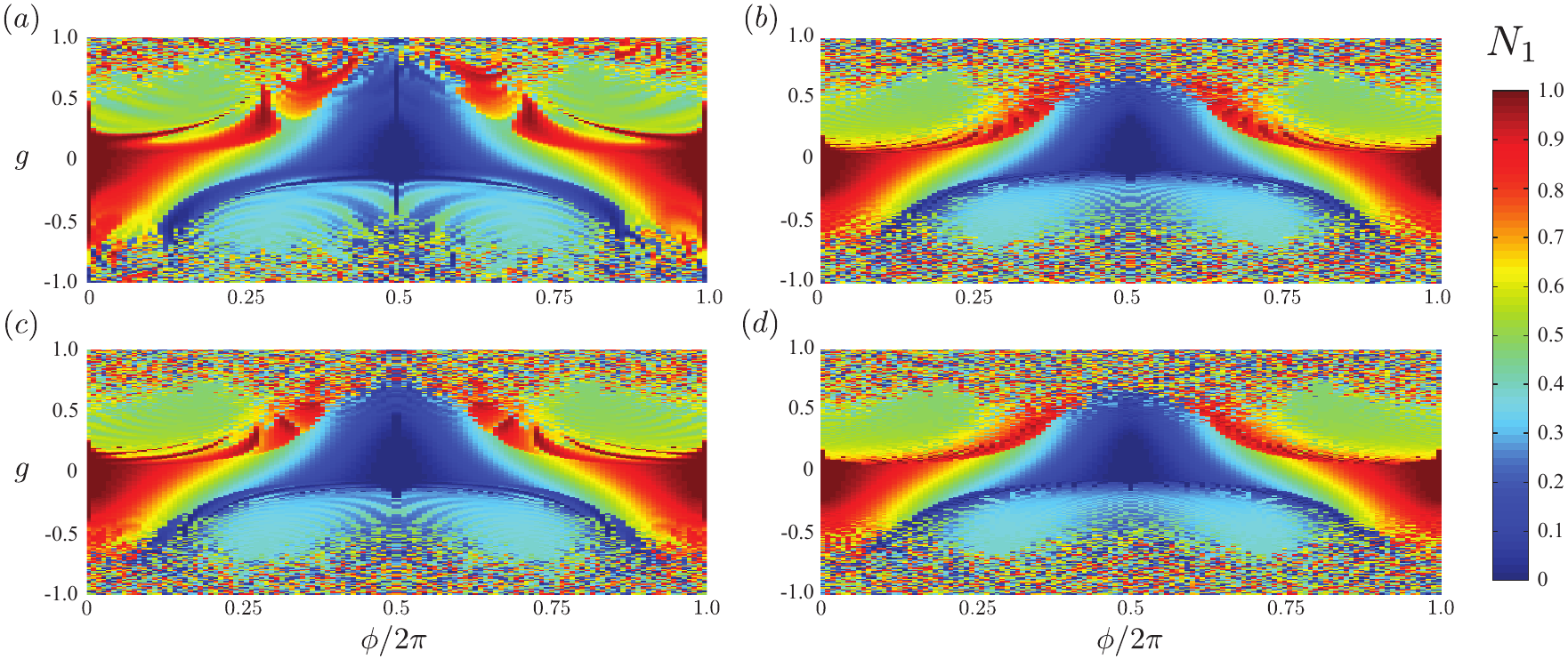}
  \end{center}
  \caption{(Color online) Occupation of well $1$ after ideal BEC splitting and recombination using reversal of f-CTAP protocol for (a) $t_p=200\Omega^{-1}$, (b)
    $t_p=500\Omega^{-1}$, (c) $t_p=1000\Omega^{-1}$, and (d) $t_p=2000\Omega^{-1}$ with $\Delta=0.1$, as a function of the phase difference ($\phi$)
    between wells $1$ and $3$ at $t=t_p/2$ and the strength of the inter-atomic interactions $|g| \le 1$.}
  \vspace{-0.0cm}
  \label{fig:Phase_carpet}
\end{figure*}
\begin{figure*}[tb!]
  \begin{center}
    \includegraphics[width=1.00\textwidth]{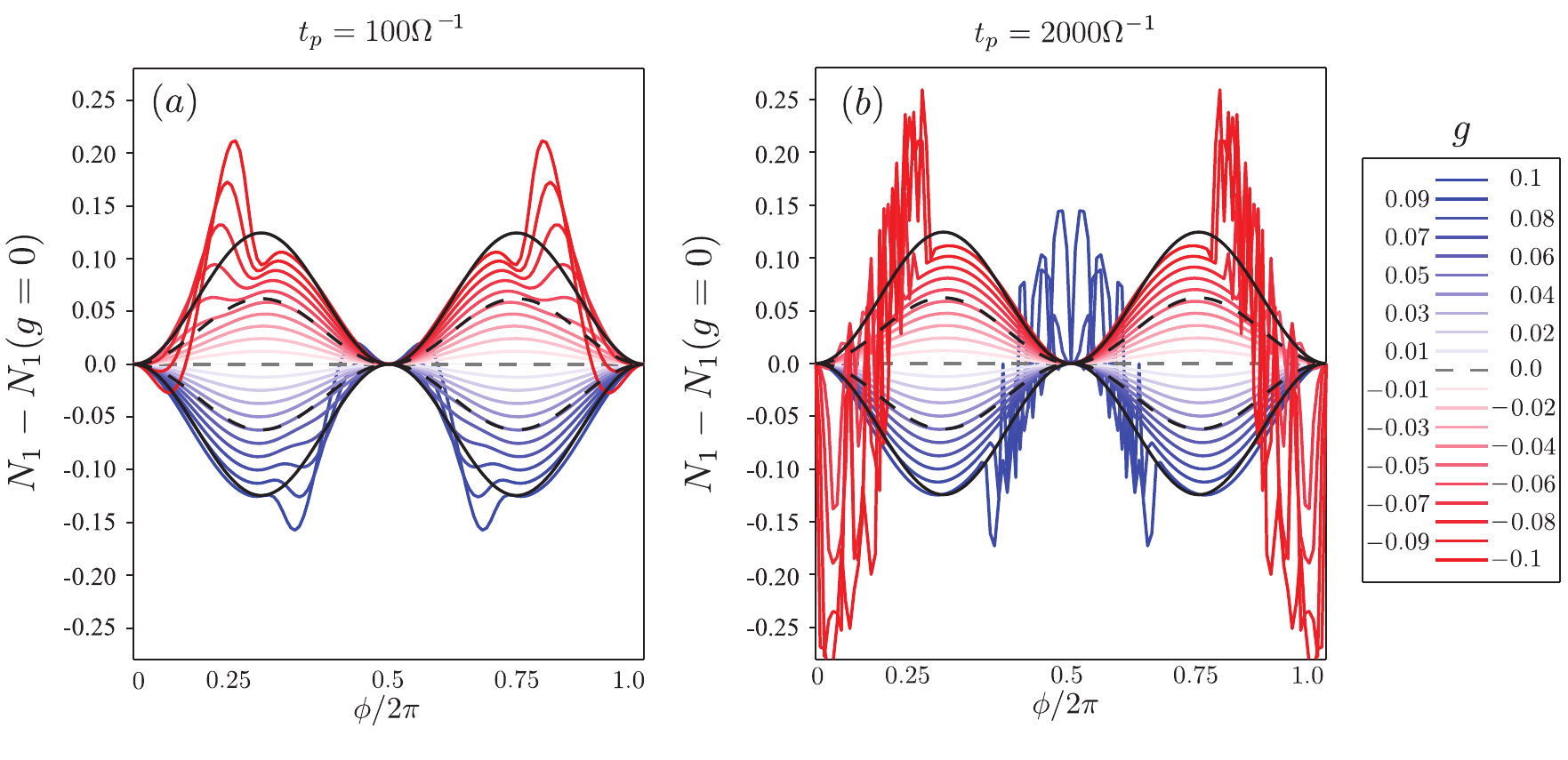}
  \end{center}
  \caption{(Color online) Deviation in the population of well $1$ from the noninteracting expected value, $N_1(|g|\le 0.1)-N_1(g=0)$, as a function of the phase
    difference ($\phi$) between wells $1$ and $3$ at $t=t_p/2$, for (a) $t_p=100\Omega^{-1}$, and (b) $t_p=2000\Omega^{-1}$, where $N_1(g=0) = \cos^2
    (\phi/2)$. The classical first order perturbation, Eq.~\eqref{equ:phidiff}, is plotted for $|g|=0.05$: dashed curves, and $|g|=0.1$: solid black
    curves.}
  \vspace{-0.0cm}
  \label{fig:FigPerturb}
\end{figure*}
As shown in Section~\ref{sec:nonintphase}, the reversal of the f-CTAP splitting process gives a phase-sensitive population in well 1,
Eq.~\eqref{ND_0}. Here we quantify the effect of interactions on this phase measurement by numerically integrating Eq.~\eqref{eq:numerical} to
determine the population of well 1 at the end of the protocol, $N_1(t_p+\tau)$. We assume an ideal splitting with
$N_1(t_p/2+\tau)=N_3(t_p/2+\tau)=0.5$, $E_1^0=E_3^0=0$ and $E_2^0=\Delta=0.1$ in Eq.~\eqref{CTAP_H}, and allow the BECs to accrue an initial relative
phase difference of $\phi$ as in Eq.~\eqref{Psi_S_three_well}.  This investigation of the phase response focuses on the behaviour for timescales
longer than the linear adiabatic pulse time where the transport is well behaved. Shorter recombination timescales result in a significant loss in
sensitivity and are not pertinent to the current investigation.\label{sec:interferometry}

\begin{figure*}[tb!]
  \centering
  \includegraphics[width=\textwidth]{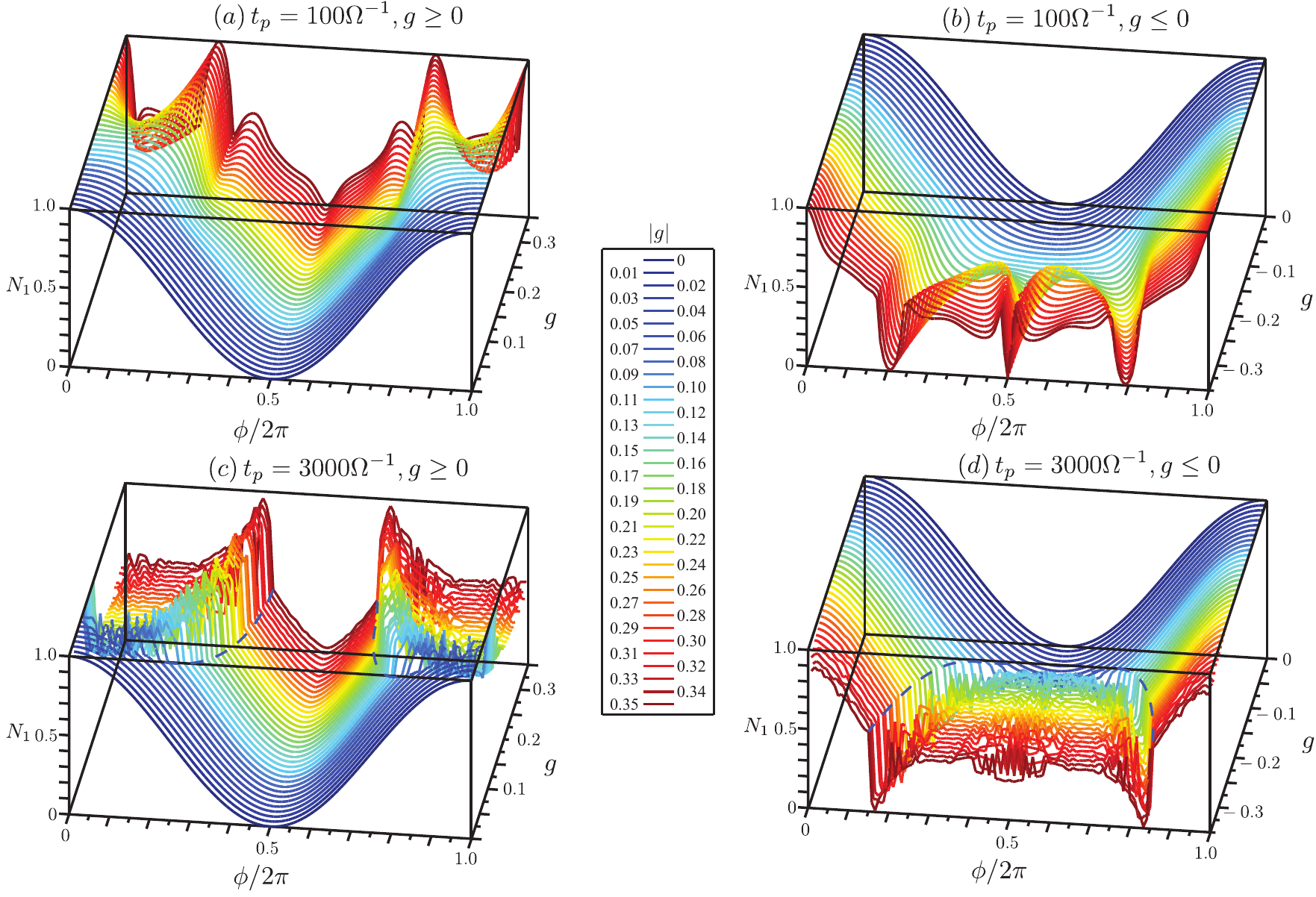}
  \caption{(Color online) Phase-sensitive measurement of the occupation of well $1$, $N_1(t_p)$, for interatomic interaction strengths $|g| \le 0.35$, after
    recombination times of $t_p = 100\Omega^{-1}$, (a) and (b), and $t_p = 3000\Omega^{-1}$ , (c) and (d). Ideal splitting of a BEC as defined in
    Eq.~\eqref{Psi_S_three_well} is assumed.}
  % \vspace{-2.0cm}
  \label{fig:tp100-3000}
\end{figure*}

The final population of well 1 for $t_p=200\Omega^{-1}$, $t_p=500\Omega^{-1}$, $t_p=1000\Omega^{-1}$ and $t_p=2000\Omega^{-1}$
(Fig.~\ref{fig:Phase_carpet}) exhibit large variations that prevent a robust phase measurement for $|g|\agt|g_c|$. For $|g|<0.2$ the phase response is
similar to the noninteracting limit [Eq.~\eqref{ND_0}] and coincides with the region of stability for the $D^{\prime}_0$ state
(Fig.~\ref{fig:lypdark}).

We have computed first-order corrections to the periodic orbits using classical perturbation theory~\cite{Goldstein2001Cl}. For small $|g|$, the
deviation of the final population well 1 from the noninteracting case, $\Delta{}N_1$, is given by:
\begin{equation} \Delta{}N_1 =N_1-N_1(g=0)= \frac{-g\Omega^2\sin^2{(\phi)}}{4\Delta(2 + \Delta^2)},
  \label{equ:phidiff}
\end{equation}
which agrees with the numerical simulations in Fig.~\ref{fig:FigPerturb}. The non-zero $\Delta{}N_1$ implies an increased phase sensitivity due to the
interactions, which is robust with respect to the pulse time.

Consistent with Eq.~\eqref{equ:phidiff}, we numerically find that $\Delta N_1$ is zero for $\phi =\{0,\pi\}$ (Fig.~\ref{fig:FigPerturb}). For long
pulse times, Fig.~\ref{fig:FigPerturb}(b), the longer time spent traversing the unstable regions of phase space leads to a larger deviation near $\phi
= \{0,\pi\}$. For small attractive interactions, Fig.~\ref{fig:FigPerturb} shows that for $\phi \approx 0$ the interferometer is stable. Conversely,
for small repulsive interactions, Fig.~\ref{fig:FigPerturb} shows that for $\phi \approx \pi$ the interferometer is stable.

Even though bifurcations do not emerge till $|g|>|g_c|\approx1.0$, Fig.~\ref{fig:Phase_carpet} demonstrates that for interaction strengths $|g|>0.5$,
the instability of the symmetric and antisymmetric stationary states at $\phi = \{0,\pi\}$ lead to large variations in density.  Large density
fluctuations reduce the possibility of recovering any phase information across all of $\phi$ for $g>0.7$.

\label{sec:enhanc-phase-sens}
\begin{figure*}[tb!]
  \centering
  % \vspace{2.0cm}
  \includegraphics[width=1.0\textwidth]{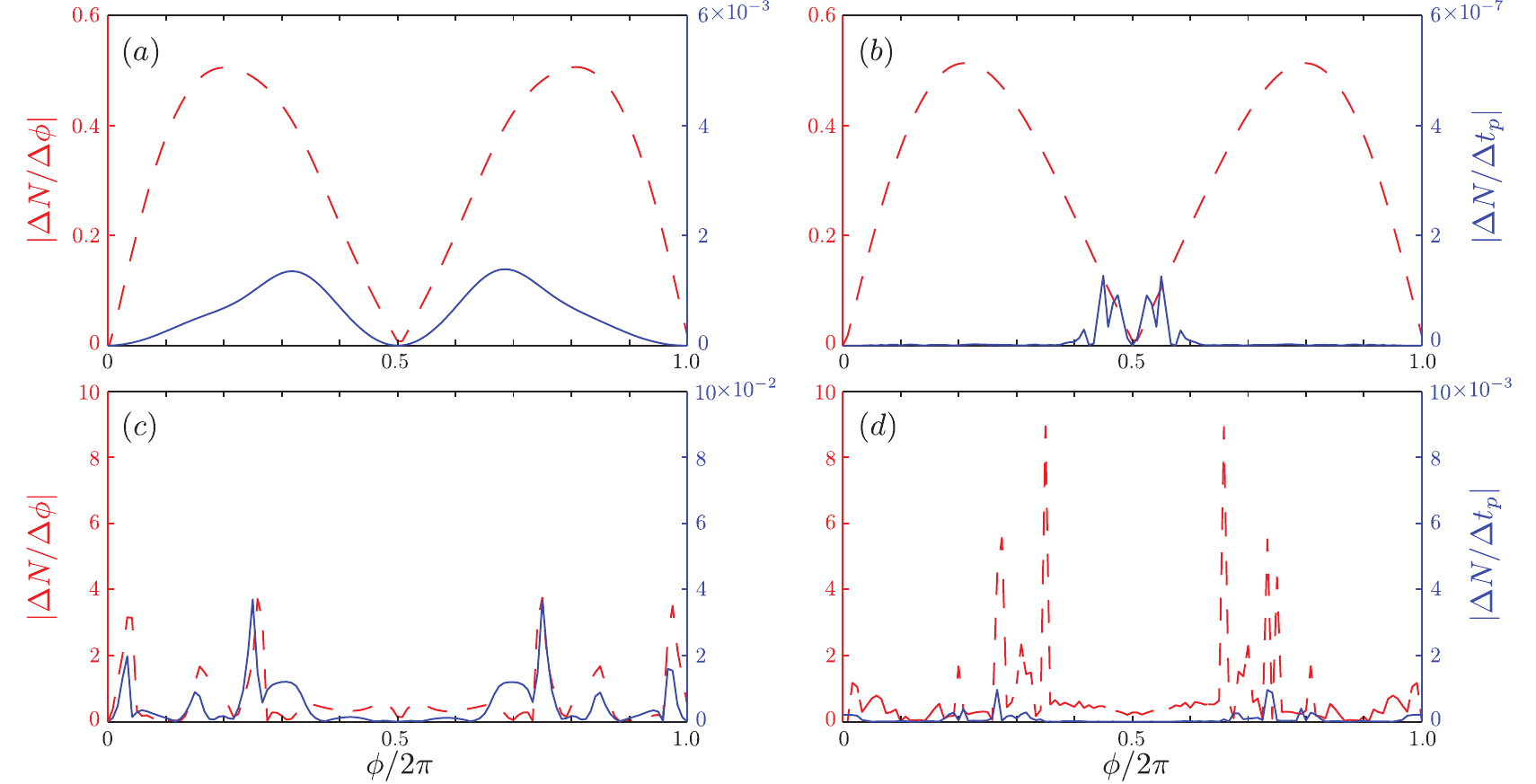}
  \caption{(Color online) Sensitivity of the density measurements for interaction strengths of (a,b) $g=-0.06$, (c,d) $g=0.3$ with respect to changes
    in phase and total pulse time. The dashed lines correspond to the phase sensitivity, $\partial N/ \partial \phi$, and the solid lines
    correspond to the pulse-time sensitivity, $\partial N/ \partial t_p$, between $90 \Omega^{-1}\le t_p\le110\Omega^{-1}$ for (a) and (c), and $1800
    \Omega^{-1}\le t_p\le2200\Omega^{-1}$ for (b) and (d).} \vspace{-0.0cm}
  \label{fig:FigNonPerturbDiff}
\end{figure*}

In Fig.~\ref{fig:tp100-3000}, we show the regions where phase-sensitive measurements can be performed in the presence of nonlinear interactions,
i.e. $|g| \alt 0.35$. Since nonlinear atom-atom interactions can introduce unwanted density-dependent fluctuations, which compromise phase
sensitivity, atomic interferometry experiments often reduce these interactions via a Feshbach resonance~\cite{Fattori2008Ma,Winkler2007Co}.

In the small interaction limit with $|g|<0.1$ a robust phase-sensitive measurement of $N_1$ can be made by accounting for the correction given by
Eq.~\eqref{equ:phidiff}. In this regime, Fig.~\ref{fig:tp100-3000} shows a smooth response of $N_1$ with respect to phase, with
deviations for the noninteracting result shown in Fig.~\ref{fig:FigPerturb}. The instability of the $\phi = \pi$ state for attractive interactions,
seen as deviation from the first-order correction in Fig.~\ref{fig:FigPerturb}, does decrease the pulse-time sensitivity relative to the stable
regions. However, the f-CTAP interferometer should be robust for a broader range of interactions.

\subsubsection{Enhanced phase sensing}

Here we highlight how a f-CTAP based interferometer has two distinct regimes of enhanced phase-sensitive operation, facilitated by control of the
interaction strength via a Feshbach resonance. Regime 1 is characterised by sensing around $\phi = \{0,\pi\}$ for short pulse times whereas regime 2
requires much longer pulse times, but allows for tuneable sensing.

In regime 1: For shorter pulse times [see Fig.~\ref{fig:Phase_carpet}(a) and \ref{fig:Phase_carpet}(b)] around $\phi = \{0,\pi\}$, an enhanced
phase-sensitive measurement of $N_1$ can be made. For small $|g|$, the first-order corrections to the final population in well 1
[Eq.~\eqref{equ:phidiff}] are small, and there is very little sensing enhancement. However, for greater $|g|$, large but stable oscillations appear in
these regions, leading to enhanced sensing. This occurs for $0.2<g<0.6$ ($\phi\approx0$) and $-0.2>g>-0.6$
($\phi\approx\pi$)~[Fig.~\ref{fig:Phase_carpet}(a)]. The emergence of these oscillations can be seen in Fig.~\ref{fig:FigPerturb}(a) where $\phi =
\{0,\pi\}$ for $g <0$ and $g >0$ respectively. As can be seen in Figs.~\ref{fig:Phase_carpet}(a)-\ref{fig:Phase_carpet}(d) the range of interaction
strengths where the $\phi = \{0,\pi\}$ states return to $N_1=\{1,0\}$ reduces with increasing pulse times.

In regime 2: A highly enhanced phase-sensitive measurement of $N_1$ can be made at the point indicated by the dashed line in
Figs.~\ref{fig:tp100-3000}(c)~and~\ref{fig:tp100-3000}(d). This stems from the large gradient in the population phase response in the vicinity of the
dashed lines. The position of this boundary could be controlled by tuning the interaction strength via a Feshbach resonance. This would, for instance,
allow the position of maximum sensitivity for a given phase to be changed depending on the application. This opens up the possibility of using a
feedback-based phase measurement to ensure maximum sensitivity to particular phase changes.

\subsubsection{Robustness}
\label{sec:robustness}

As shown in Section~\ref{sec:stability}, the adiabaticity criteria, Eq.~\eqref{equ:adiabaticcondition}, is invalidated by the introduction of the
nonlinear interaction term. Hence, here we associate robust phase-sensitive measurement with a high sensitivity to changes in relative phase,
$\partial N/ \partial\phi$, and a low sensitivity to changes in pulse time, $\partial N/ \partial t_p$.

For the small interaction limit, where the divergence due to the instability is relatively slow,
Figs.~\ref{fig:FigNonPerturbDiff}(a)~and~\ref{fig:FigNonPerturbDiff}(b) shows that as we increase the pulse time from $t_p = 100\Omega^{-1}$ to $t_p =
2000\Omega^{-1}$ the pulse-time sensitivity decreases by a factor of $10^{-4}$. However, this deviation is small with respect to phase
sensitivity. While this regime exhibits minimal enhancement of phase sensitivity, it will have the robustness and stability of a noninteracting f-CTAP
interferometer.

In regime 1: There is some reduction in pulse-time sensitivity relative to the small interaction limit [Fig.~\ref{fig:FigNonPerturbDiff}], however the
pulse-time sensitivity is still a factor of 10 less than the phase sensitivity for $t_p= 100\Omega^{-1}$ and is further reduced by a factor of $10^3$
for $t_p = 2000\Omega^{-1}$ [Figs.~\ref{fig:FigNonPerturbDiff}(c) and \ref{fig:FigNonPerturbDiff}(d)]. The nonlinear interaction also leads to a
smaller enhancement of the phase-measurement for $g<0.3$ and $g>0.3$ for $\phi = \{0,\pi\}$, which is more robust to changes in pulse time due to the
reduced instability. Despite this, Fig.~\ref{fig:FigNonPerturbDiff}(d) shows that the phase enhancement persists for longer pulse times.

In regime 2: Surprisingly, the robustness of sensing for this range of phase shifts is increased for longer pulse times where peaks in phase
sensitivity no longer align with peaks in pulse-time sensitivity [Fig.~\ref{fig:FigNonPerturbDiff}(d)]. The presence of regions in the parameter space
that are insensitive to total pulse time implies that robust interferometry is possible for the interacting BEC.

\section{Conclusion}
We have demonstrated that the f-CTAP protocol is a valid methodology for the coherent spatial
splitting of a BEC in the presence of interactions. Reversal of the f-CTAP protocol provides a
robust phase-sensitive measurement as an alternative to traditional methods. This interferometric
process is robust to changes in pulse time once in the adiabatic regime for the small interaction
limit. Also, even though the adiabatic principle is not strictly valid in this nonlinear system, it
is possible to find splitting and recombination timescales that achieve the balance between
maintaining the system in the $D^{\prime}_0$ stationary state and divergence due to
instability. Phase-sensitive measurements in the presence of nonlinear atomic interactions can lead
to an enhancement in sensitivity without significant loss in robustness with respect to changes in
pulse time.

\section*{Acknowledgements}
The authors thank Charles Hill for helpful discussions. A.D.G.~acknowledges the Australian Research Council for
financial support (Project No.~DP0880466).
\bibliography{bib}
\end{document}